Kleptoparasitism on carpenter ants (*Camponotus* spp.) by *Podarcis tiliguerta* (GMELIN, 1789) in Corsica and *Podarcis filfolensis* (BEDRIAGA, 1876) on the Maltese Islands

In June 2009, a foraging trail of the carpenter ant *Camponotus vagus* was observed in a pine forest area near Calacuccia, central north Corsica. The weather was bright and sunny at approximately 18 °C. *Podarcis tiliguerta* (GMELIN, 1789) were active. As the ants collected a number of dead and live invertebrates and plant particles, an adult *P. tiliguerta* was observed approaching the ant trail. For approximately two minutes the lizard watched and moved up and down the ant trail at a distance of about one meter. The lizard then spotted three ants dragging a cricket. The lizard approached the trail and swiftly grabbed the orthopteran, shaking it and hitting it on the ground, subsequently removing two of the ants; it then retreated fast into nearby vegetation and was observed to consume the insect.

Kleptoparasitic behavior was also observed by the authors in *P. filfolensis* (BEDRIAGA, 1876) in a Maltese Island. In August 2006 on the unpopulated Cominotto Island, a female *P. filfolensis* was observed feeding on a trail of *Camponotus barbaricus* ants. Ants comprise approximately 16 % of the lizard's diet on this island during summer (SCIBERRAS unpublished data). A trail of *C. barbaricus* just 2.5 m ahead of the aforementioned *P. filfolensis* carried a dead *Sphingonotus coerulans* locust. The lizard stopped feeding on the ants and collected the dead *S. coerulans*, rushing under the closest thicket of *Thymbra capitata* and then rested in the shade to consume the prey. On a second occasion, a population of *C. barbaricus* disturbed by human trampling were relocating their eggs, food seeds of *Ferula communis*, and undetermined seeds from one location toward a small rubble wall. Several specimens of *P. filfolensis* were observed robbing eggs and shaking off the ants.

During mating season, this lacertid generally attends strict territorial boundaries and rarely forages outside of them. Territories vary from 1.2 m² to 2.7 m² (BORG 1989; SCIBERRAS unpublished data). Interestingly, ants were also observed carrying items on their trail with lizards attending, but not attacking, the colony. This could possibly be because the activity was assembled outside of *P. filfolensis* territory.

Kleptoparasitism (parasitism by theft) is a form of feeding in which one animal takes prey or other food from another that has caught, collected, or otherwise prepared the food, including stored food. The behavior, may it be intra- or interspecific, is present in a number of animals; mostly, insects, birds and humans (COYL et al. 1991; JORDE & LINGLE 1998; SIVINSKI et al. 1999; SHEALER et al. 2005; SCHOE et al. 2009). It is rarely seen in wild lizards but has been reported for another Mediterranean island lacertid, *Podarcis lilfordi* (GÜNTHER, 1874) (COOPER & PÉREZ-MELLADO 2003).

Kleptoparasitism on ants is more rarely observed in temperate and mediterranean ecosystems, in part due to lower densities of ants and generally smaller sized ant prey, which would reduce the energetic benefit gained by the parasite. Kleptopararistic behavior on ants is more common in the tropics where swarms of army ants (*Eciton burchellii* in the Neotropics and *Dorylus molestus* in Africa - SCHÖNING et al. 2006), generate numerous examples of kleptoparasitism due to their considerable biomass and predictable behavior. Kleptoparasites of these examples include specialized guilds of birds (e.g., BROCKMANN & BARNARD 1979) that follow ant swarms and glean invertebrates, lizards or snakes that try and escape them.

That this behavior has been observed in two mediterranean species of *Podarcis*, independently parasitizing on species of *Camponotus* ant, suggests there is benefit to kleptoparasitism on *Camponotus* spp. in Mediterranean ecosystems. The benefits of this example of kleptoparasitism may partly be derived from the large size of the ant (workers of *C. vagus* are typically 6-12 mm, whilst those of *C. barbaricus* are typically 10-15 mm) which attract predators such as *Podarcis*, initially to feed directly on them, but also to indulge in opportunistic kleptoparasitism on larger insect prey caught by the ants. BROCKMANN & BARNARD (1979) suggest that occurrences of kleptoparasitism are more frequent during situations of food shortage as can seasonally occur on Mediterranean islands. As species of *Campono­tus* are global in their distribution (YAMAMOTO & DEL-CLARO 2008), it is possible that incidences of kleptoparasitism by other species of lizard are currently unrecorded.

The benefits and costs that determine the profitability of food stealing by lizards likely depend on intrinsic characteristics of individuals that facilitate or constrain kleptoparasitic behavior. It has been hypothesized that kleptoparasitism should be more profitable, and hence have more often evolved, in taxonomic lineages featuring certain characteristics, such as a large body mass, an enlarged brain or a dependence on certain prey types. Alternatively, the evolution of kleptoparasitism in lizards could have been facilitated in certain ecological contexts, such as open habitats or mixed-species foraging groups (GIRALDEAU & CARACO 2000; COOPER & PÉREZ-MELLADO 2003) or by the predictability of ant species on which the behavior will yield a high food return for low effort.

From the observations above, it is evident that both *P. filfolensis* and *P. tiliguerta* are interspecific kleptoparasites of *Campo­notus* spp. ants.

ACKNOWLEDGMENTS: The authors thank Anne LEONARD, Colin RYALL, Tim JENKINS and Pete RENDALL of the registered charity Operation New World < www.opnewworld.co.uk > for support and the opportunity to study the ecology of Corsica's herpetofauna. Jeffrey SCIBERRAS and Esther SCIBERRAS are acknowledged for their continuous assistance with the field work on the Maltese islands. Jordan WAGENKNECHT is thanked for identification of *Camponotus vagus* from Corsica.

AUTHORS: Todd R. LEWIS (Corresponding author < ecolewis@gmail.com >) [1], Alex RAMSAY [2], Arnold SCIBERRAS [3] and Colin BAILEY [4]

[1] Westfield, 4 Worgret Road, Wareham, Dorset, BH20 4PJ, UK

[2] Calls Wharf, 2 The Calls, Leeds, LS2 7JU, UK

[3] 133 Arnest, Arcade Str., Paola, Malta

[4] Truffles, Piggery Hall Lane, West Wittering, Chichester, West Sussex, PO20 8PZ, UK